\documentclass{llncs}
\usepackage{proof}
\newcommand{\defeq}{\mathrel{\stackrel{{\scriptstyle\triangle}}{=}}}
\newcommand{\Dscr}{{\cal D}}
\newcommand{\false}{\hbox{\sl false}}

\begin{document}
\title{The Bedwyr system for model checking over syntactic expressions}
\author{
        David Baelde\inst{1} \and
        Andrew Gacek\inst{2} \and 
        Dale Miller\inst{1} \and
        Gopalan Nadathur\inst{2} \and \\
        Alwen Tiu\inst{3}
       }
\institute{
  INRIA \& LIX, \'Ecole Polytechnique \and
  Digital Technology Center and Dept of CS, University of Minnesota\and
  Australian National University and NICTA
}
\maketitle

\section{Overview}

Bedwyr is a generalization of logic programming that allows model
checking directly on syntactic expressions possibly containing
bindings. This system, written in OCaml, is a direct implementation of
two recent advances in the theory of proof search.  The first is
centered on the fact that both finite success and finite failure can
be captured in the sequent calculus by incorporating inference rules
for {\em definitions} that allow {\em fixed points} to be explored.
As a result, proof search in such a sequent calculus can capture
simple model checking problems as well as may and must behavior in
operational semantics.  The second is that higher-order abstract
syntax is directly supported using term-level $\lambda$-binders and
the $\nabla$ quantifier.  These features allow reasoning
directly on expressions 
containing bound variables.

\section{Foundations}

The logical foundation of Bedwyr is the logic called LINC
\cite{tiu04phd}, an acronym for ``lambda, induction, nabla, and
co-induction'' that is an enumeration of its major components.  LINC
extends intuitionistic logic in two directions.

\paragraph{Fixed points via definitions.}
Clauses such as $A\defeq B$ are used to provide (mutually) recursive 
definitions of atoms. Once a set $\Dscr$ of such definition clauses
has been fixed, LINC provides inference rules for introducing atomic
formulas based on the idea of unfolding definitions. Unfolding on the
right of the sequent arrow is specified by the following {\em
  definition-right} rule:
\[
\vcenter{
  \infer
      {\Sigma:\Gamma\vdash A}
      {\Sigma:\Gamma\vdash B\theta}}
  \hbox{\ , provided $A'\defeq B\in\Dscr$ and $A'\theta=A$.}
\]
This rule resembles backchaining in more conventional logic
programming languages.  The {\em definition-left} rule is a case analysis
justified by a closed-world reading of a definition.
\[
\infer{\Sigma:\Gamma, A \vdash G}
      {\{\Sigma\theta:\Gamma\theta,B\theta\vdash G\theta
       \ | \ A'\defeq B\in\Dscr\mbox{ and }
       \theta \in csu(A, A')\}}
\]
Notice that this rule uses unification: the eigenvariables of
the sequent (stored in the signature $\Sigma$) are instantiated by
$\theta$, which is a member of a complete set of unifiers (csu) for
atoms $A$ and $A'$.  Bedwyr implements a subset of this rule that is
restricted to {\em higher-order pattern unification} and, hence, to
a case where {\em csu} can be replaced by {\em mgu}.  If an atom on
the left fails to unify with the head of any definition, the premise
set of this inference rule is empty and, hence, the sequent is proved:
thus, a unification {\em failure} is turned into a proof search {\em
success}. 

Notice that this use of definitions as fixed points implies that logic
specifications are not treated as part of a {\em theory} from which
conclusions are drawn.  Instead, the proof system itself is
parametrized by the logic specification.  In this way, definitions
remain fixed during proof search and the {\em closed world assumption}
can be applied to the logic specification.  For earlier references to
this approach to fixed points see
\cite{girard92mail,schroeder-heister93lics,mcdowell97lics}.

\paragraph{Nabla quantification.}
Bedwyr supports the \emph{$\lambda$-tree syntax} \cite{miller00cl}
approach to higher-order abstract syntax \cite{pfenning88pldi} 
by implementing a logic that provides
$(i)$ terms that may contain $\lambda$-bindings, $(ii)$ variables that
can range over such terms, and $(iii)$~equality (and unification) that
follows the rules of $\lambda$-conversion.  Bedwyr shares these
attributes with systems such as $\lambda$Prolog. However, it
additionally includes the $\nabla$-quantifier that is needed to fully
exploit the closed-world aspects of LINC. This quantifier can be read
informally as ``for a new variable'' and is accommodated easily
within the sequent calculus with the introduction of a new kind of
local context scoped over formulas.  We refer the reader to
\cite{miller05tocl} for more details.  We point out here, however,
that $\nabla$ can always be given minimal scope by using the
equivalences $\nabla x.(A x * B x) \equiv (\nabla x. A
x) * (\nabla x.  B x)$ where $*$ may be $\supset$, $\land$ or $\lor$ and
the fact that $\nabla$ is self-dual: $\nabla x.\lnot Bx 
\equiv\lnot \nabla x. Bx$.  When $\nabla$ is moved under $\forall$ and
$\exists$, it {\em raises} the type of the quantified variable: in
particular, in the equivalences 
$\nabla x\forall y. F x y \equiv\forall h\nabla x. F x (h x)$ and 
$\nabla x\exists y. F x y \equiv\exists h\nabla x. F x (h x)$,
the variable $y$ is replaced with a functional variable $h$.  Finally,
when $\nabla$ is scoped over equations, the equivalence
$\nabla x(T x = S x) \equiv (\lambda x.T x)=(\lambda x.S x)$
allows it to be completely removed. As a result, no fundamentally new
ideas are needed to implement $\nabla$ in a framework where
$\lambda$-term equality is supported.

\section{Architecture}

Bedwyr implements a fragment of LINC that is large enough to permit interesting 
applications of fixed points and $\nabla$. 
In this fragment, {\em all} the left rules are 
invertible.  Consequently, we can use a simple proof strategy 
that alternates between left and right-rules, with the left-rules 
taking precedence over the right rules.

{\em Two provers.} The fragment of LINC implemented in Bedwyr is given 
by the following grammar: 
\begin{eqnarray*}
L0 &::=&
  \top\; |\; A\; |\; L0 \land L0\; |\; L0 \lor L0\; |\; \nabla x.~ L0\; |\; \exists x.~ L0 \\
L1 &::=& 
  \top\; |\; A\; |\; L1 \land L1\; |\; L1 \lor L1\; |\; \nabla x.~ L1\; |\; \exists x.~ L1\; |\;
  \forall x.~ L1\; |\; L0 \supset L1 \\
\end{eqnarray*} 
The formulas in this fragment are divided into {\em level-0} formulas, 
given by $L0$ above, and {\em level-1} formulas, given by $L1$. 
Implicit in the above grammar is the partition
of atoms into level-0 atoms and level-1 atoms. Restrictions apply to goal 
formulas and definitions: goal formulas can be level-0 
or level-1 formulas, and in a definition $A \defeq B$, $A$ and $B$ can
be level-0 or level-1 formulas, provided that the level of $A$
is greater than or equal to the level of $B$. 

Level-0 formulas are essentially a subset of 
goal formulas in $\lambda$Prolog (with $\nabla$ replacing $\forall$).
Proof search for a defined atom of level-0 is thus
the same as in $\lambda$Prolog (and Bedwyr implements that fragment
following the basic ideas described in \cite{elliott91semi}).
We can think of a level-0 definition, 
say, $p\,x \defeq B\,x$, as defining a set of elements $x$ 
satisfying $B\,x$. A successful proof search for $p\,t$ means 
that $t$ is in the set characterized by $B$. 
A level-1 statement like $\forall x. p\,x \supset R\,x$ would then
mean that $R$ holds for all elements of the set characterized by $p$.
That is, this statement captures the enumeration of a {\em model}
of $p$ and its verification can be seen as a form of model checking. 
To reflect this operational reading of level-1 implications,
the proof search engine of Bedwyr uses two subprovers: 
the Level-0 prover (a simplified $\lambda$Prolog engine), 
and the Level-1 prover.
The latter is a usual depth-first goal-directed prover but with a
novel treatment of implication.  
When the Level-1 prover reaches the implication $A\supset B$, it
calls the Level-0 prover on $A$ and gets in return a stream of answer
substitutions: the Level-1 prover then checks that, for every 
substitution $\theta$ in that stream, $B\theta$ holds.
In particular, if Level-0 finitely fails with $A$, the implication is proved. 

As with most depth-first implementations of proof search, Bedwyr
suffers from some aspects of incompleteness: for example, the prover
can easily loop during a search although different choices of goal or
clause ordering can lead to a proof, and certain kinds of
unification problems should be delayed instead of attempted eagerly.
For a more detailed account on the incompleteness issues, we refer the
reader to \cite{tiu05eshol}. 
Bedwyr does not currently implement static checking of types and the
stratification of definitions (which is required in the
cut-elimination proof for LINC). This allows us to experiment with a
wider range of examples than those allowed by LINC.

{\em Higher-order pattern unification.}  We adapt the treatment of
higher-order pattern unification due to Nadathur and
Linnell~\cite{nadathur05iclp}. This implementation uses the {\em
suspension calculus} representation of $\lambda$-terms.  We avoid
explicit raising, which is expensive, by representing $\nabla$-bound
variables by indices and associating a {\em global} and a {\em local}
level annotation with other quantified variables.  
The global level replaces raising over existential and universal 
variables.
The local level replaces raising over $\nabla$-bound variables.
For example, the scoping in
$\forall{}x.\exists{}y.\nabla{}n.\forall{}z. F x y n z$
is represented by the following annotation:
$F x^{0,0} Y^{1,0} \#_0 z^{2,1}$
(we use lowercase letters for universal variables,
uppercase for existentials,
the index $\#_n$ for the $n$-th $\nabla$-bound variable,
and write in superscript the annotation $(global,local)$).
Using this annotation scheme, the scoping aspects of $\nabla$ quantifiers
are reflected into new conditions on local levels
but the overall structure of the higher-order pattern unification
problem and its mgu properties are preserved.  

{\em Tabling.} We introduced tabling in Bedwyr to cut-down exponential 
blowups caused by redundant computations and to detect loops during 
proof-search. The first optimization is critical for applications such 
as weak bisimulation checking. The second one proves useful when exploring 
reachability in a cyclic graph. 

Tabling is currently used in Bedwyr to experiment with proof search for
inductive and co-inductive predicates. A loop over an inductive predicate 
that would otherwise cause a divergence can be categorized using
tabling as a failure. 
Similarly, in the co-inductive case, loops yield success. 
This interpretation of loops as failure or success is not part of the
meta-theory of LINC. Its soundness is currently conjectured, although
we do not see any inconsistency of this interpretation on the numerous
examples that we tried. 

Inductive proof-search with tabling is implemented effectively
in provers like XSB~\cite{sagonas06xsb} using, for example, 
suspensions.  The implementation of tables in Bedwyr  
fits simply in the initial design of the prover but is much weaker.
We only table a goal 
in Level-1 when it does not have free occurrences of variables
introduced by an existential quantifier; and 
in Level-0 when it does not have any free variable occurrence.
Nevertheless, this implementation of tabling has proved useful 
in several cases, ranging from graph examples to bisimulation.

\newcommand{\step}[2]{\hbox{\sl step}~#1~#2}
\newcommand{\win }[1]{\hbox{\sl win} ~#1}
\newcommand{\simm}[2]{\hbox{\sl sim}~#1~#2}
\newcommand{\one }[3]{#1\stackrel{#2}{-\!\!-\!\!\!\to    } #3}
\newcommand{\onep}[3]{#1\stackrel{#2}{-\!\!-\!\!\!\rightharpoonup} #3}

\section{Examples}

We give here a brief description of the range of applications of Bedwyr.
We refer the reader to {\tt\url{http://slimmer.gforge.inria.fr/bedwyr}} 
and the user manual for Bedwyr \cite{baelde06manual} for more details about
these and other examples.

{\em Finite failure}.  Let $\false$ be an atom that
has no definition.  Negation of a level-0 formula $G$ can then be
written as the level-1 formula $G\supset\false$ and this negation is
provable in the level-1 prover if all attempts to prove $G$ in the
level-0 prover fail. For example, the formula $\forall y [\lambda
x.x = \lambda x.y \supset \false]$ is a theorem: {\em i.e.}, the
identity abstraction is always different from a constant-valued
abstraction.

{\em Model-checking}.
If the two predicates $P$ and $Q$ are defined using Horn clauses, then
the Level-1 prover is capable of attempting a proof of
$\forall x.~ P~x\supset Q~x$.
This covers most (un)reachability checks common in model-checking.
Related examples in the Bedwyr distribution include the verification of a 3 bits 
addition circuit and graph cyclicity checks.

{\em Games and strategies.}
Assuming that a transition in a game from position $P$ to position
$P'$ can be described by a level-0 formula $\step{P}{P'}$ then
proving the level-1 atom $\win P$ defined by
\[ 
 \win P\defeq\forall P'.~\step P{P'}\supset\exists P''.~\step{P'}{P''}\land\win{P''}
\]
will determine if there is a winning strategy from position $P$.  If
all {\sl win}-atoms are tabled during proof search, the resulting table
contains an actual winning strategy.

{\em Simulation in process calculi.}  
If the level-0 atom $\one{P}{A}{Q}$ specifies a one-step
transition (process $P$ does an $A$-action and results in process
$Q$), then simulation can be written in Bedwyr as follows \cite{mcdowell03tcs}. 
\[ 
 \simm P Q\defeq\forall A\forall P'.
    ~\one{P}{A}{P'}\supset\exists Q'.\;\one{Q}{A}{Q'}\land\simm{P'}{Q'}
\]
In dealing with the $\pi$-calculus, where bindings can occur within
one-step transitions, there are two additional transitions that need
to be encoded: in particular, $\onep{P}{\downarrow X}{P'}$ and
$\onep{P}{\uparrow X}{P'}$, for bound input and bound output
transitions on channel $X$.  In both of these cases, $P$ is a process
but $P'$ is a name abstraction over a process.  The full specification
of (late, open) simulation for the $\pi$-calculus can be written using
the following \cite{miller05tocl}.
\[
\simm{P}{Q}\defeq
\begin{array}[t]{l}
 [\forall A\forall P'
    \begin{array}[t]{l}
      .\;\one{P}{A}{P'}\supset\exists Q'.\;\one{Q}{A}{Q'}
       \land \simm{P'}{Q'}]\land\null
    \end{array} \\ \relax
 [\forall X\forall P'
        \begin{array}[t]{l}
        .\;\onep{P}{\downarrow X}{P'}\supset
         \exists Q'.\;\onep{Q}{\downarrow X}{Q'}
          \land \forall w.\simm{(P'w)}{(Q' w)}]\land \null          
    \end{array} \\ \relax   
 [\forall X \forall P'
    \begin{array}[t]{l}
      .\;\onep{P}{\uparrow X}{P'}\supset\exists Q'.\;\onep{Q}{\uparrow X}{Q'}
         \null\land\nabla w. \simm{(P' w)}{(Q' w)}]
    \end{array}
\end{array}
\]
Notice that the abstracted continuation resulting from bound input and
bound output actions are treated by the $\forall$-quantifier and the
$\nabla$-quantifier, respectively.  In a similar way, modal logics for
the $\pi$-calculus can be captured \cite{tiu05concur}.
If {\sl sim}-atoms are tabled during proof search, the resulting table
contains an actual simulation.  Bisimulation is easily captured by
simply adding the symmetric clauses for all those used to define {\sl sim}.

{\em Meta-level reasoning}.  
Because Bedwyr uses the $\nabla$ quantifier and the $\lambda$-tree approach to encoding syntax, 
it is possible to specify provability in an object logic
and to reason to some extent about what is and is not provable.  
Consider the tiny fragment of intuitionistic logic with the universal 
quantifier $\overline \forall$ and the implication $\Rightarrow$ in which
we only allow atoms to the left of implications. 
If the formula 
$\overline\forall x.~(p~x~r\Rightarrow \overline\forall y.~(p~y~s\Rightarrow p~x~t))$
is provable in this logic then one would expect $r$ and $t$ to be syntactically
equal terms. In searching for a proof of this formula, the quantified variables are
replaced by distinct eigenvariables: therefore, the only way the formula could 
have been proved is for $p~x~t$ to match $p~x~r$, hence $r = t.$
Provability of a formula $B$ from a list of atomic formulas $L$ 
can be specified by the following meta-level (Bedwyr-level) judgment $pv~L~B$:
\[
\begin{array}{l}
pv~L~B \defeq memb~B~L. \qquad\qquad\qquad
pv~L~(\overline\forall B) \defeq \nabla x.~pv~L~(B\,x).\\
pv~L~(A \Rightarrow B) \defeq pv~(A::L)~B.
\end{array}
\]
Here, $memb$ and $::$ are the usual predicate for list membership and 
the non-empty list constructor. Object-level eigenvariables are
specified using the meta-level $\nabla$-quantifier. 
The above observation about object-logic provability can now be stated
in the meta-logic as the following formula, which is provable in Bedwyr:
$$\forall r\forall s \forall t.~
  pv~nil~(\overline\forall x.~(p~x~r\Rightarrow
          \overline\forall y.~(p~y~s\Rightarrow p~x~t)))
\supset r = t.$$

\section{Future Work}

We are working on several improvements to Bedwyr, including more
sophisticated tabling and allowing the suspension of goals containing 
non-higher-order-pattern unification (rescheduling them when
instantiations change them into higher-order pattern goals).
We will also explore using tables as proof certificates: for example,
when proving that two processes are bisimilar, the table stores an
actual bisimulation, the existence of which proves the bisimilarity.
Bedwyr is an open source project: more details about it can be found at
{\tt \url{http://slimmer.gforge.inria.fr/bedwyr/}}.

\paragraph{Acknowledgments.} 
Support has been obtained for this work from the following
sources: from INRIA through the ``Equipes Associ{\'e}es'' Slimmer,
from the ACI grant GEOCAL, from the NSF Grants OISE-0553462
(IRES-REUSSI) and CCR-0429572 and from a grant from Boston Scientific.

\bibliographystyle{plain}

\begin{thebibliography}{10}

\bibitem{baelde06manual}
David Baelde, Andrew Gacek, Dale Miller, Gopalan Nadathur, and Alwen Tiu.
\newblock {\em A User Guide to {Bedwyr}}, November 2006.

\bibitem{elliott91semi}
Conal Elliott and Frank Pfenning.
\newblock A semi-functional implementation of a higher-order logic programming
  language.
\newblock In Peter Lee, editor, {\em Topics in Advanced Language
  Implementation}, pages 289--325. MIT Press, 1991.

\bibitem{girard92mail}
Jean-Yves Girard.
\newblock A fixpoint theorem in linear logic.
\newblock An email posting to the mailing list linear@cs.stanford.edu, February
  1992.

\bibitem{mcdowell97lics}
Raymond McDowell and Dale Miller.
\newblock A logic for reasoning with higher-order abstract syntax.
\newblock In {\em Proc. LICS 1997}, pp. 434--445, IEEE Comp. Soc. 
  Press, 1997.

\bibitem{mcdowell03tcs}
Raymond McDowell, Dale Miller, and Catuscia Palamidessi.
\newblock Encoding transition systems in sequent calculus.
\newblock {\em Theoretical Computer Science}, 294(3):411--437, 2003.

\bibitem{miller00cl}
Dale Miller.
\newblock Abstract syntax for variable binders: An overview.
\newblock In John Lloyd and {et. al.}, editors, {\em Computational Logic - {CL}
  2000}, number 1861 in LNAI, pages 239--253. Springer, 2000.

\bibitem{miller05tocl}
Dale Miller and Alwen Tiu.
\newblock A proof theory for generic judgments.
\newblock {\em ACM Trans.\ on Computational Logic}, 6(4):749--783, October
  2005.

\bibitem{nadathur05iclp}
Gopalan Nadathur and Natalie Linnell.
\newblock Practical higher-order pattern unification with on-the-fly raising.
\newblock In {\em {ICLP 2005: 21st International Logic Programming
  Conference}}, volume 3668 of {\em LNCS}, pages 371--386, Sitges, Spain,
  2005. Springer.

\bibitem{pfenning88pldi}
Frank Pfenning and Conal Elliott.
\newblock Higher-order abstract syntax.
\newblock In {\em Proceedings of the {ACM}-{SIGPLAN} Conference on Programming
  Language Design and Implementation}, pages 199--208. ACM Press, June 1988.

\bibitem{sagonas06xsb}
Konstantinos Sagonas, Terrance Swift, David~S. Warren, Juliana Freire, Prasad
  Rao, Baoqiu Cui, Ernie Johnson, Luis de~Castro, Rui~F. Marques, Steve Dawson,
  and Michael Kifer.
\newblock {\em The {XSB} Version 3.0 Volume 1: Programmer's Manual}, 2006.


\bibitem{schroeder-heister93lics}
Peter Schroeder-Heister.
\newblock Rules of definitional reflection.
\newblock In {\em Proc. LICS 1993}, pages 222--232. IEEE Comp. Soc. Press, 1993.

\bibitem{tiu04phd}
Alwen Tiu.
\newblock {\em A Logical Framework for Reasoning about Logical Specifications}.
\newblock PhD thesis, Pennsylvania State University, May 2004.

\bibitem{tiu05concur}
Alwen Tiu.
\newblock Model checking for $\pi$-calculus using proof search.
\newblock In M. Abadi and L. de~Alfaro, editors, {\em CONCUR},
  volume 3653 of {\em LNCS}, pages 36--50. Springer, 2005.

\bibitem{tiu05eshol}
Alwen Tiu, Gopalan Nadathur, and Dale Miller.
\newblock Mixing finite success and finite failure in an automated prover.
\newblock In {\em Proc. of ESHOL'05: Empirically Successful Automated
  Reasoning in Higher-Order Logics}, pages 79 -- 98, December 2005.

\end{thebibliography}

\end{document}